\documentclass[
reprint,
aps,
pra,
superscriptaddress
]{revtex4-2}
\usepackage{comment}
\usepackage{graphicx}
\usepackage[margin=1in]{geometry}
\usepackage{float}
\usepackage{comment}
\usepackage{tikz}
\usetikzlibrary{positioning}
\usepackage{amsmath}
\usepackage{pstricks,pst-all}

\usepackage{textcomp}
\usepackage{gensymb}
\usepackage{caption}
\usepackage{subcaption}
\usepackage{wrapfig}
\usepackage{graphicx}
\usepackage{mathtools}
\usepackage{tabularx}
\usepackage{amssymb}
\usepackage[normalem]{ulem}
\usepackage{amsmath, nccmath}
\usepackage{hyperref}
\usepackage{commath}
\usepackage{amsthm}
\usepackage{placeins}
\usepackage{bm}

\usepackage{lineno}
\usepackage{footmisc}
\makeatletter
\newcommand*{\rom}[1]{\expandafter\@slowromancap\romannumeral #1@}
\makeatother

\usepackage{orcidlink} 
\usepackage{lipsum}
\makeatletter
\def\ps@pprintTitle{%
 \let\@oddhead\@empty
 \let\@evenhead\@empty
 \def\@oddfoot{}%
 \let\@evenfoot\@oddfoot}
\makeatother
\begin{document}

\title{Study of Randomness in Quantum Random Number Generator using Bayesian Neural Networks}

\author{Shriram Pragash M\,\orcidlink{0009-0003-2110-1118}\textsuperscript{\textdagger}}
\affiliation{Indian Institute of Technology Hyderabad, Telangana, Sangareddy - 502284, India.}
\email{yvabhignan@gmail.com, venkat.ph@nitp.ac.in}
\author{Aditi Das\,\orcidlink{0009-0005-9668-1991}\textsuperscript{\textdagger}}
\affiliation{Qdit Labs Pvt. Ltd., Bengaluru - 560092, India.}
\email{These authors contributed equally.}
\author{R. Srikanth\,\orcidlink{0000-0001-7581-2546}}
\affiliation{Poornaprajna Institute of Scientific Research, Bengaluru - 562110, India.}

\author{Megha Shrivastava\,\orcidlink{0000-0001-5000-7970}}
\affiliation{Qdit Labs Pvt. Ltd., Bengaluru - 560092, India.}

\author{Mohit Mittal}
\affiliation{Qdit Labs Pvt. Ltd., Bengaluru - 560092, India.}

\author{Venkat Abhignan\,\orcidlink{0000-0002-6092-3670}\textsuperscript{\textdagger,$*$}}
\affiliation{Qdit Labs Pvt. Ltd., Bengaluru - 560092, India.}
    \affiliation{Department of Applied Physics and Materials Engineering, National Institute of Technology Patna, Bihar - 800005, India}

\begin{abstract} We propose a Bayesian neural network (BNN) framework to quantify the intrinsic (aleatoric) uncertainty in a continuous-variable quantum random number generator (CV-QRNG). This provides a direct data-driven measure of the QRNG’s randomness generation. The BNN is trained on raw homodyne voltage time-series data from a fiber-coupled vacuum-fluctuation QRNG (VFQRNG). For each voltage data segment, the network predicts a Gaussian distribution (mean and variance) for the next sample, and the ensemble-averaged predicted variance is used as a diagnostic of entropy-related behavior under assumptions of our quantum source model. Thus, the predicted variance serves as a BNN model-based monitor of the randomness-generation process. We validate that the BNN metric increases with the local-oscillator (LO) power, consistent with theoretical min-entropy (extractable randomness). We also show the metric is sensitive to practical imperfections: coarse-graining (simulating limited analog-to-digital converter (ADC) sampling) or reducing ADC bit depth causes a clear drop in predicted aleatoric uncertainty, consistent with the source min-entropic model. These results establish the BNN-based metric as a complementary, physically motivated diagnostic tool for QRNG characterization. Unlike conventional tests (e.g. autocorrelation) that miss hardware issues like low ADC resolution, our method provides a direct real-time monitor of the VFRNG performance. Therefore, this metric is useful for real-time monitoring of CV-QRNG conditions, including drifts and other imperfections. \end{abstract}

\maketitle
\section{INTRODUCTION}
\label{sec:intro}
Random numbers are essential in science, cryptography, and simulations \cite{RevModPhys.89.015004}. Based on deterministic techniques, pseudo-RNGs \cite{10.5555/270146} are commonly used to generate random numbers \cite{hörmann2004automatic}. Nevertheless, given sufficient computational resources, their output can be shown to exhibit intrinsic periodicity, as it depends only on the original seed and the particular method used. For applications like cryptography, where predictability undermines security, intrinsic randomness is essential.

QRNGs exploit intrinsic quantum randomness \cite{stipvcevic2011quantum,Ma2016}. Most single-photon-detection-based QRNGs have been limited to a comparatively low random number sampling rate due to low detector count rates ($<$100 MHz) \cite{SINGLE1,SINGLE2,SINGLE5}. Significantly higher rates are achieved by homodyne measurement \cite{collett} of vacuum quadratures \cite{Yuen:83}. This uses coherent detection using balanced homodyne photodetectors, eliminating the dead-time bottleneck of single-photon detection \cite{Gabriel2010,10.1063/1.3597793,PhysRevApplied.3.054004}. In addition, Toeplitz randomness extractor implementations on FPGAs can output random bits at Gb/s \cite{10.1063/1.5078547}. Thus, these high-speed systems require robust methods to characterize randomness directly on the analog signal.

QRNG security improves under source-device-independent (SI) assumptions \cite{PhysRevLett.118.060503,PhysRevApplied.12.034017}. Recently, large-scale SI-QRNGs on silicon photonic chips now reach 20–200 Gb/s \cite{Huang:19,10.1088/2058-9565/ae4d81}. We have also implemented a fiber-coupled SI-QRNG \cite{Shrivastava2025}.  This ensures security even with an untrusted source. However, measurement devices can remain vulnerable to security threats. Measurement-device-independent (MDI) QRNG schemes aim to eliminate all possible manipulation on the measurement apparatus by leveraging entanglement/Bell inequality violations \cite{Pironio2010, Nie2016}. However, such MDI QRNGs are single–photon–based with significant technical complexity due to entanglement generation, synchronization, and low detector count rates. While SI is achieved in CV-QRNGs, MDI for homodyne systems is still unproven. For entanglement-based QRNGs, public-testing schemes have been proposed in which one of multipartite entanglement-coupled bitstreams is disclosed for verification while the remaining streams are kept secret \cite{Jacak2020,Jóźwiak2024}. Although this enables extensive testing of long-range patterns, it relies on a dedicated multipartite entanglement architecture and is not directly transferable to a CV-QRNG. For CV-SI-QRNGs, a power spectral density (PSD)-based conditional-variance framework was recently proposed to account for detector-bandwidth-induced temporal correlations and derive a lower bound on min-entropy against quantum side information \cite{Zhou2026}.

In homodyne CV-QRNGs, extractable randomness depends on the detection system parameters. Specifically, the vacuum quadrature variance (and theoretical min-entropy) increases with LO power until detectors saturate \cite{Gabriel2010,PhysRevApplied.3.054004,10898923}. Consequently, scanning LO power calibrates the quantum-to-classical noise ratio: higher LO power amplifies vacuum noise over detector classical noise, usually increasing extractable entropy \cite{Zhang2021,PhysRevA.109.052401,PhysRevA.102.012422,Shrivastava2025}. From a security perspective, altering the LO or injecting signals can bias measurements and reduce true randomness, as shown experimentally  \cite{PhysRevA.100.052318,Zhou_2022,PhysRevApplied.15.044044}.  

Studies show that although entropy bounds depend on LO power, simple diagnostics (variance or autocorrelation function (ACF)) are insensitive to those changes once detector bandwidth effects are considered \cite{PhysRevA.87.062327,Kim2023,Zhang2021}. Detector nonidealities (finite bandwidth) and ADC effects (finite range, quantization) can greatly reduce min-entropy without showing up as long-range correlations in ACF tests \cite{Zhang2021,PhysRevA.109.052401,PhysRevA.102.012422}. For instance, modeling ADC discretization and bandwidth shows that undersampling or coarse quantization suppresses high-frequency fluctuations, causing significant underestimation of min-entropy if ignored \cite{Zhang2021,PhysRevA.102.012422,PhysRevA.109.052401}. Thus, despite LO sweeps being useful, a diagnostic gap remains: conventional metrics (variance, ACF) struggle to detect randomness degradation in the raw signal, motivating more sensitive data-driven analysis.

Neural networks have become powerful tools for quantum analysis \cite{RevModPhys.91.045002}, including QRNGs \cite{9187201,Crespo_2024}. Typically, machine learning techniques are employed to evaluate the randomness of bit streams after classical post-processing and randomness extraction \cite{9187201,Crespo_2024}, and these works found predictions fail when quantum noise dominates, confirming the QRNG’s robustness. However, these methods rely on reference distributions and do not directly quantify the raw signal’s behaviour. Moreover, these methods operate on post-processed bit streams, where classical extraction and digitization effects may obscure the underlying quantum source behavior.

Unlike device-independent protocols \cite{Pironio2010, Nie2016,Vallone2014,Liu2018,Zhou_2024}, CV protocols rely on trusted physical models to estimate extractable min-entropy $H_{\text{min}}$ \cite{Gabriel2010,PhysRevApplied.3.054004,10898923,Zhang2021,PhysRevA.109.052401,PhysRevA.102.012422,Shrivastava2025}. However, unmodelled correlations or drifts can compromise randomness. We propose a complementary, data-driven monitor (a min-entropy model-free diagnostic) using a BNN framework to characterize the randomness generation process in a CV homodyne QRNG from its raw analog voltage output. Unlike prior studies on post-processed bits \cite{9187201,Heese2024,Crespo_2024} our approach operates directly on the homodyne output and provides a data-driven assessment of the fluctuations contributing to the randomness-generation process, while remaining sensitive to detector noise and practical imperfections. We validate this on a experimental fiber-coupled vacuum-homodyne QRNG, showing the BNN uncertainty metric behaves as expected with changing parameters.

We train an ensemble of BNNs on segments of the homodyne output voltage, where each network is tasked with predicting the next voltage sample in a segment and outputs both a mean and a variance. The mean captures predictable structure, while the predicted variance $U$ represents the BNN model-inferred uncertainty (aleatoric uncertainty). As LO power increases (increasing vacuum variance $\sigma_Q$ over classical noise $\sigma_E$) \cite{Gabriel2010,PhysRevApplied.3.054004,10898923}, we observe the average predicted uncertainty $\overline{U}$ rises accordingly, mirroring the expected increase in min-entropy $H_{\min}(\sigma_Q)$. Finally, we show the method is sensitive to ADC imperfections \cite{Zhang2021,PhysRevA.109.052401,PhysRevA.102.012422}, such as reduced resolution (using coarse-graining $\eta$) and limited sampling rate $n$, highlighting its use as a real-time monitor of QRNG performance. We emphasize that this is a complementary diagnostic, not a replacement for formal entropy certification with adversarial considerations, which require explicit assumptions about the source model, adversarial side information, and detector characterization. It does not guarantee unconditional security against adversarial side-information.

The remainder of the paper is organized as follows. Sec.~\ref{sec:setup} describes the fiber-coupled homodyne-based CV-QRNG which measures the vacuum-state quadrature fluctuations. Sec.~\ref{sec:laser-power} and the subsequent Secs.~\ref{sec:adc1},~\ref{sec:adc2} present the evaluation of the proposed BNN uncertainty metric under the considered operating LO power and realistic ADC conditions, respectively with the corresponding comparisons to conventional ACF analysis. The main findings, scope, limitations, and implications of the approach are discussed in Sec.~\ref{summary}. Supplement~A provides the detailed data-preparation and BNN motivation along with methodology, including time-series voltage data segmentation, normalization, input-output construction, network architecture, discussions on variational inference and predicted aleatoric variance $U$. Supplement~B gives the mathematical definitions used for the ACF and for modeling ADC finite sampling and quantization. Supplement~C reports NIST SP~800-22 statistical test of the generated random bit data from the VFQRNG.

\section{EXPERIMENTAL SETUP}
\label{sec:setup}

\begin{figure*}[tbp]
    \centering
\includegraphics[width=0.75\textwidth]{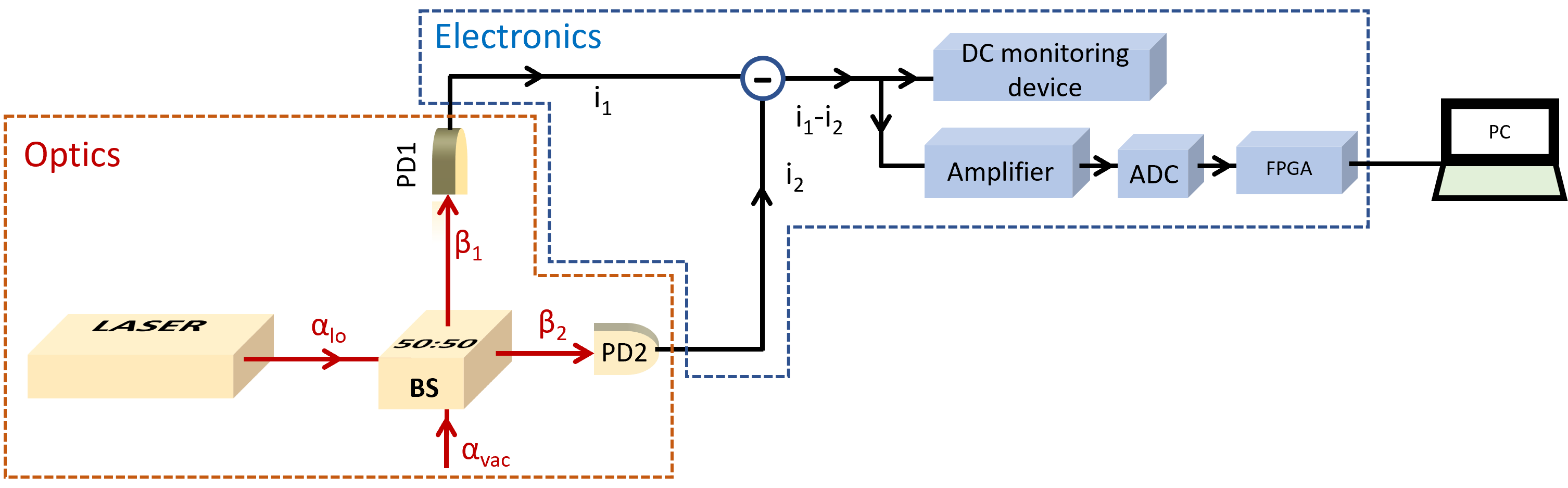}
    \caption{Schematic diagram of the experimental setup used.}
    \label{fig:vfqrng}
\end{figure*}

Our experimental configuration consists of a completely fiber-coupled assembly for the randomness generation. As shown in Fig. \ref{fig:vfqrng}, the interference of vacuum state ($\alpha_{vac}$) with a strong LO ($\alpha_{lo}$) (TeraXion PS-NLL-1550) at a 50:50 beam splitter (50:50 BS), yields two outputs $\beta_1$ and $\beta_2$. Calibrated photodetectors measure each output and subtract their currents ($\text{i}_{12} = \text{i}_1-\text{i}_2$). The AC component of $\text{i}_{12}$ is the vacuum-fluctuation signal. This signal is amplified, monitored, and digitized by an FPGA ADC. The balanced detector used is a Thorlabs PDB480C-AC (1.6 GHz) 

The homodyne output voltage is $\propto \left< \text{i}^2_{12} \right> - \left< \text{i}_{12} \right>^2$. The sampled raw data of this voltage consists of the vacuum signal and has a normal distribution with variance $\sigma_V$. However, classical noise from the detectors invariably taints the raw output of voltage. This classical side-information-independent randomness conditioned on this classical noise can be measured by min-entropy \cite{PhysRevApplied.3.054004,10.1063/1.5078547} \begin{equation}
    H_{\text{min}}=-\log_2\left[\text{erf}\left(\frac{\delta}{2\sqrt{2}\sigma_{Q}}\right)\right],
\end{equation}
depending on vacuum signal variance $\sigma_Q$ and quantization error $\delta$ of the 14-bit ADC implemented. This shows that $H_{\text{min}}$ random bits can be obtained by utilising ($H_{\text{min}}$/.14)\% of the raw voltage data. Here, the variance of the vacuum signal $\sigma_Q=\sqrt{\sigma_V^2-\sigma_E^2-2(\delta/n)^2}$ can be obtained from the total variance $\sigma_V$, and the variance $\sigma_{E}$ is from the sampled raw voltage data, which exhibits a non-zero value when the laser is turned off. This non-zero $\sigma_{E}$ is ascribed to classical noise arising from electromagnetic disturbances, temperature fluctuations, and intrinsic imperfections within the experimental configuration. ADC quantization error $\delta=2R/(2^n)$ is influenced by the dynamic voltage range $R=1$V, and the sampling precision is $n=14$ of the ADC used. 
 \begin{figure}[tbp]
    \centering
\includegraphics[width=7cm]{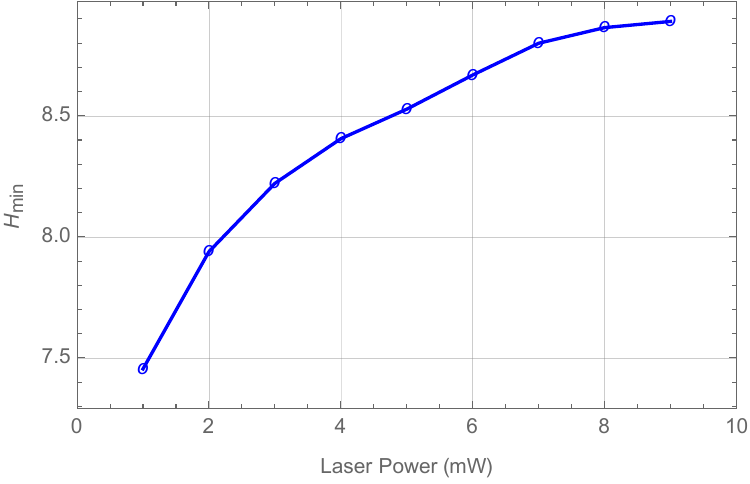}
    \caption{$H_{\text{min}}$ vs LO power. The $H_{\text{min}}$ obtained from variance of quantum signal in sampled raw voltage data.}
    \label{fig:Hmin}
\end{figure}

The optimal performance for achieving maximum randomness, as measured by $H_{\text{min}}$, is achieved by adjusting the laser's power. The variance of the voltage $\sigma_V$ measured at homodyne detectors increases linearly with laser power, ranging from 0 mW to 9 mW. Saturation is attained at a laser power of 9 mW, as can be seen in the corresponding $H_{\text{min}}$ in Fig. \ref{fig:Hmin}. This implies a maximum of 8.89 random bits can be extracted from 8.89/.14 = 64\% of the raw voltage data at 9 mW.
\section{Bayesian Neural network RESULTS}
\subsection{Analysis of Laser Power Variation}
\label{sec:laser-power}
\begin{figure*}[tbp]
    \centering
    \begin{subfigure}[b]{0.59\textwidth}
        \includegraphics[width=\textwidth]{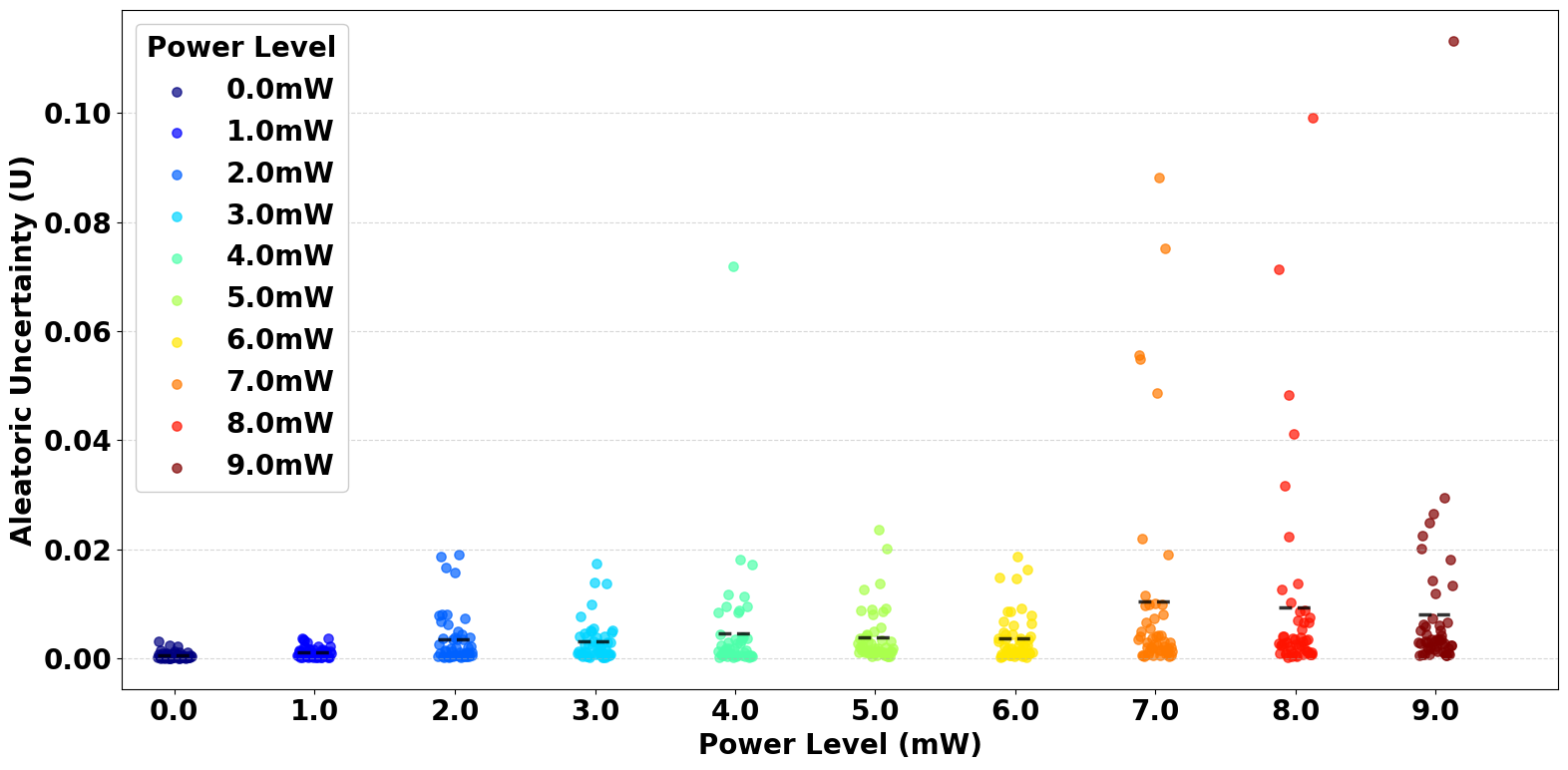}
        \caption{}
    \end{subfigure}
    \hfill
    \begin{subfigure}[b]{0.4\textwidth}
        \includegraphics[width=\textwidth]{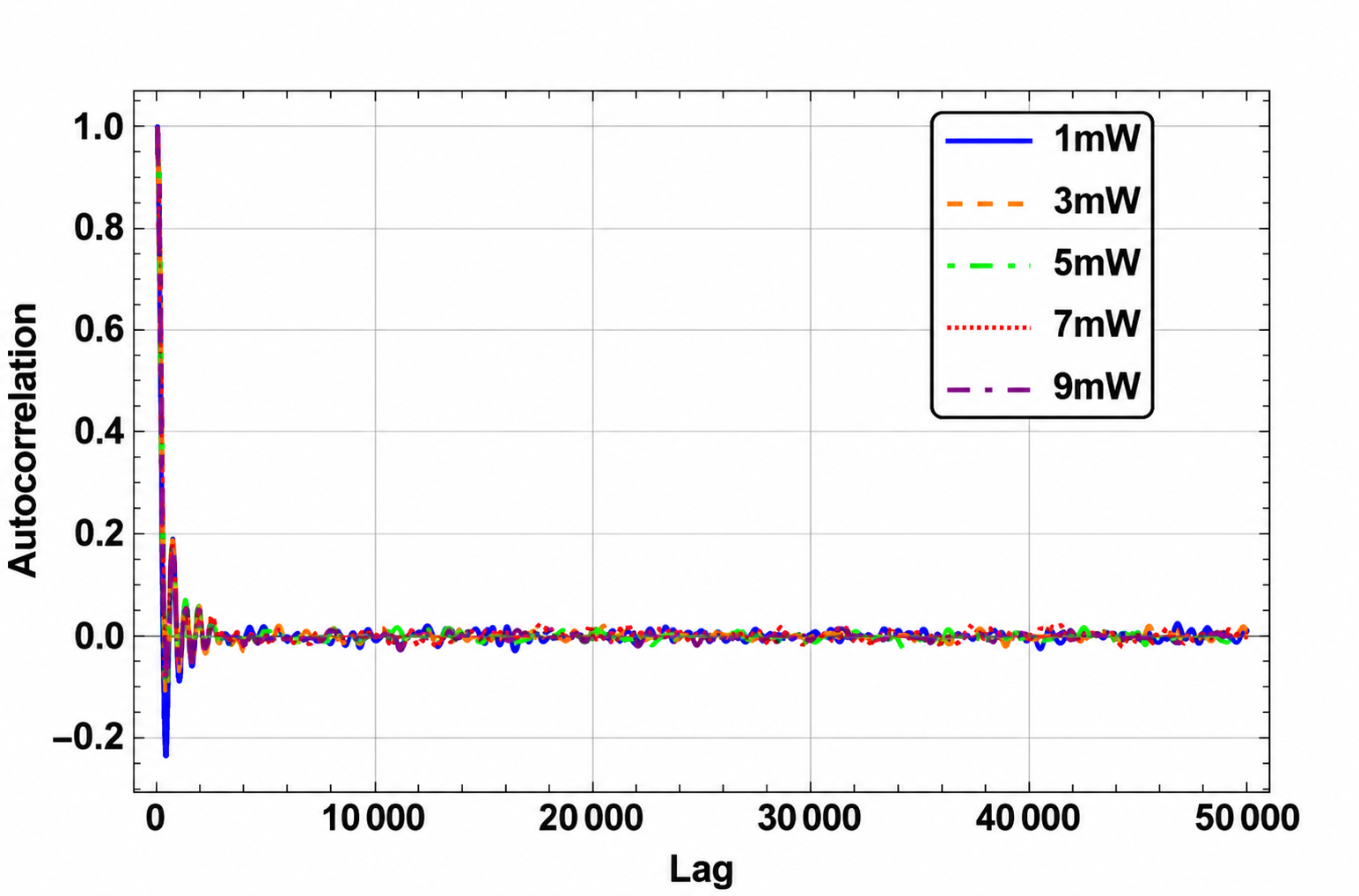}
        \caption{}
    \end{subfigure}
    \caption{(a) Aleatoric Uncertainty $U$ vs Laser power level in mW (The black dashed line indicates the mean $\overline{U}$). 
             (b) Autocorrelation of voltage data from homodyne detections at different laser powers. 
             }
    \label{fig:laser_power_analysis}
\end{figure*}

Raw homodyne voltage data acquired at LO powers of 1–9 mW were used to compute the BNN aleatoric variance $U$ (BNN evaluation described in Supplement A). 
As described in Sec.~\ref{sec:intro}, LO-power variation provides a controlled operating parameter for the homodyne QRNG \cite{Zhang2021,PhysRevA.102.012422,PhysRevA.109.052401}; here, we examine its effect on the neural-network uncertainty metric $U$. As shown in Fig. \ref{fig:laser_power_analysis} (a) (the $U$ values) and in Table \ref{tab:stddev_mse_laser} (the mean $\overline{U}$ and standard deviation $\sigma_{U}$), the ensemble-averaged uncertainty $\overline{U}$ generally increases and its standard deviation $\sigma_U$ shows variability with increasing LO power, although the trend is not strictly monotonic at every power level. This qualitative trend is consistent with the increase in source-model $H_{\text{min}}$ with increasing LO power shown in Fig. \ref{fig:Hmin}. The observed increase in BNN metric $U$ is qualitatively consistent with the expectation that increasing LO power increases the vacuum-noise contribution relative to classical detector noise, thereby increasing QCNR and the source-model $H_{\text{min}}$ (or extractable randomness) \cite{PhysRevA.102.012422,PhysRevApplied.3.054004}. However, this comparison does not imply that $U$ itself is an estimator or bound for $H_{\text{min}}$.
	

\begin{table}[ht]
\centering
\begin{tabular}{|c|c|c|}
\hline
LO Power (mW) & $\overline{U}$ & $\sigma_{U}$ \\
\hline
0  & 5.326$\times 10^{-4}$ & 6.392$\times 10^{-4}$ \\
1  & 1.126$\times 10^{-3}$ & 9.702$\times 10^{-4}$ \\
2  & 3.464$\times 10^{-3}$ & 4.745$\times 10^{-3}$ \\
3  & 3.025$\times 10^{-3}$ & 3.674$\times 10^{-3}$ \\
4  & 4.642$\times 10^{-3}$ & 1.052$\times 10^{-2}$ \\
5  & 3.839$\times 10^{-3}$ & 4.842$\times 10^{-3}$ \\
6  & 3.747$\times 10^{-3}$ & 4.371$\times 10^{-3}$ \\
7  & 1.038$\times 10^{-2}$ & 1.937$\times 10^{-2}$ \\
8  & 9.349$\times 10^{-3}$ & 1.849$\times 10^{-2}$ \\ 
9  & 8.140$\times 10^{-3}$ & 1.668$\times 10^{-2}$ \\ 
\hline
\end{tabular}
\caption{Mean aleatoric uncertainty ($\overline{U}$) and its standard deviation ($\sigma_{U}$) at different LO power levels.}
\label{tab:stddev_mse_laser}
\end{table}
We compare the BNN metric with the ACF (mathematical description given in Supplement B) \cite{PhysRevA.87.062327,Kim2023} and plot the ACF of the voltage signal for different LO power levels in Fig. \ref{fig:laser_power_analysis} (b). As observed, the function decays rapidly to zero, though initial damped oscillations (ringing) are visible at low lags ($<5,000$). The low-lag ringing is consistent with the finite bandwidth of the implemented 1.6-GHz homodyne detector. At larger lags, no appreciable LO power-dependent difference is observed within the analyzed range. Thus, the ACF does not reveal a corresponding LO power-dependent change in second-order linear correlation within the analyzed lag range. In contrast, the BNN uncertainty metric $U$ shows a LO power-dependent change that is not evident in the ACF. These results indicate that the BNN provides a complementary diagnostic of LO power operating-condition-dependent changes in the measured voltage process that are not resolved by the ACF alone \cite{PhysRevA.109.052401,Zhang2021,PhysRevA.102.012422}.

\subsection{Studying imperfect sampling effects of ADC}
\label{sec:adc1}

We use coarse-graining (block-averaging) to emulate reduced effective sampling in the homodyne data, following the existing approaches used to study finite-sampling effects in CV-QRNGs \cite{Zhang2021,PhysRevA.109.052401,PhysRevA.102.012422}. The impact of imperfect ADC sampling has been previously investigated in the context of pulsed lasers \cite{PhysRevA.109.052401}, where it was observed that increasing the sampling frequency leads to a higher value of $H_{\text{min}}$. In our work, we aim to study analogous effects in a continuous-wave laser setup. Reduced effective sampling can suppress high-frequency components of the measured voltage fluctuations. We therefore apply block averaging at retention fractions $\eta=0.1,0.3,0.5,0.7,0.8$ on the homodyne voltage data (mathematical procedure described in Supplement B).

The non-overlapping block average acts as a rectangular averaging filter followed by decimation. Previous CW homodyne and security analyses have used analogous decimation or filtering procedures to model finite analog bandwidth and bound correlations when estimating $H_{\text{min}}$ \cite{Zhang2021,PhysRevA.102.012422}. This transformation effectively simulates the loss of high-frequency voltage fluctuations due to the ADC's finite sampling bandwidth. A larger coarse-graining percentage $\eta \times 100 \%$ retains more data, corresponding to a higher effective sampling frequency of the ADC.

    \begin{table}[ht]
    \centering
    \begin{tabular}{|c|c|c|c|}
    \hline
    \begin{tabular}{c}
          Coarse-graining \\
        ($\eta \times 100 \%$) 
    \end{tabular}
     & $\overline{U}$ & $\sigma_{U}$  \\
    \hline
    10 & $7.135 \times 10^{-4}$ & $2.439 \times 10^{-4}$ \\
    30 & $1.682 \times 10^{-3}$ & $1.323 \times 10^{-3}$ \\
    50 & $1.631 \times 10^{-3}$ & $1.643 \times 10^{-3}$ \\
    70 & $7.837 \times 10^{-3}$ & $1.619 \times 10^{-2}$ \\
    80 & $9.628 \times 10^{-3}$ & $1.409 \times 10^{-2}$ \\

    \hline
    \end{tabular}
    \caption{Mean Aleatoric uncertainty ($\overline{U}$) and its standard deviation ($\sigma_{U}$) at different coarse-graining percentage.}
    \label{tab:stddev_mse_cg}
\end{table}

\begin{figure*}[tbp]
    \centering
    \begin{subfigure}[b]{0.59\textwidth}
        \includegraphics[width=\textwidth]{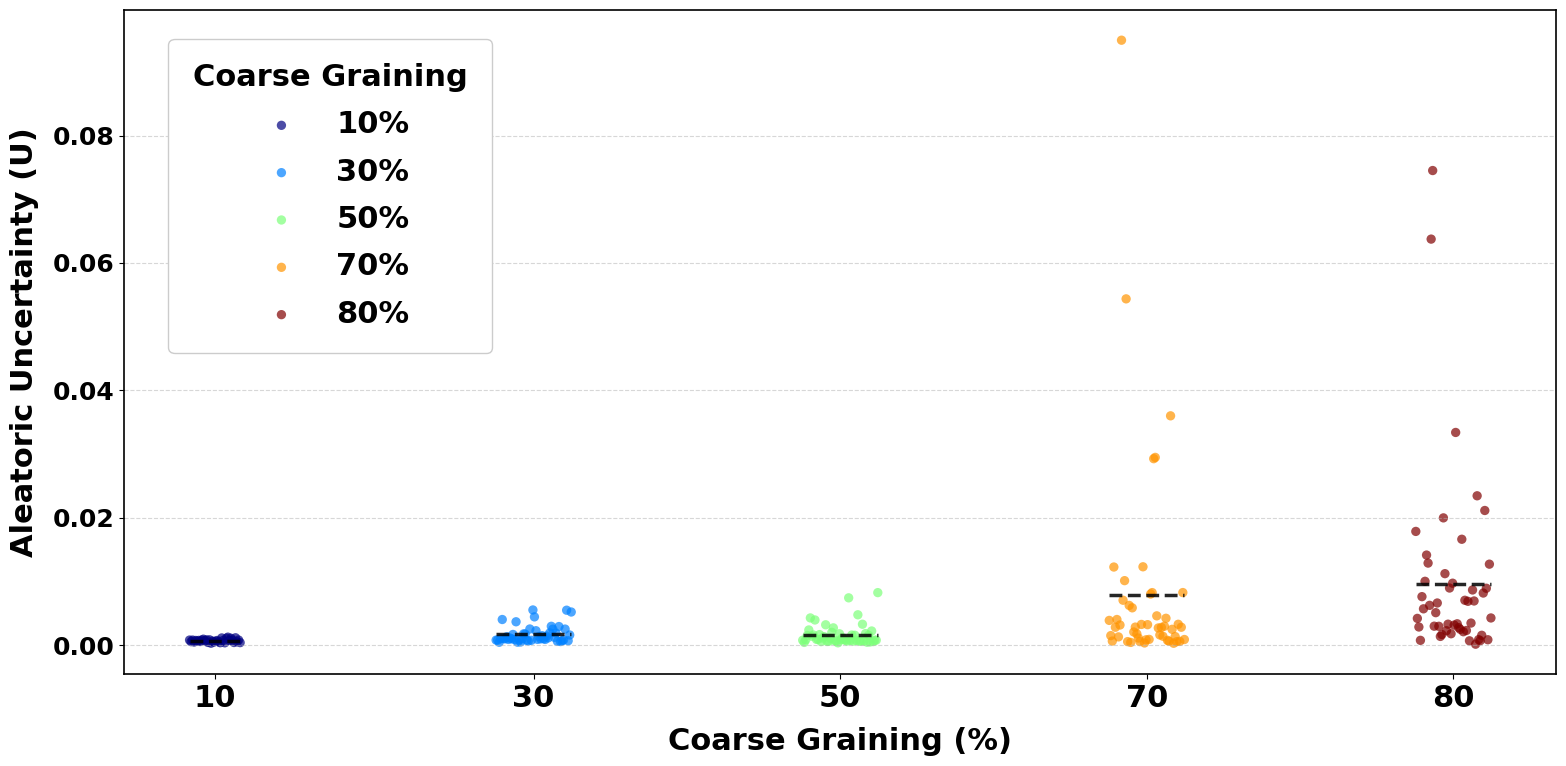}
        \caption{}
    \end{subfigure}
    \hfill
    \begin{subfigure}[b]{0.4\textwidth}
        \includegraphics[width=\textwidth]{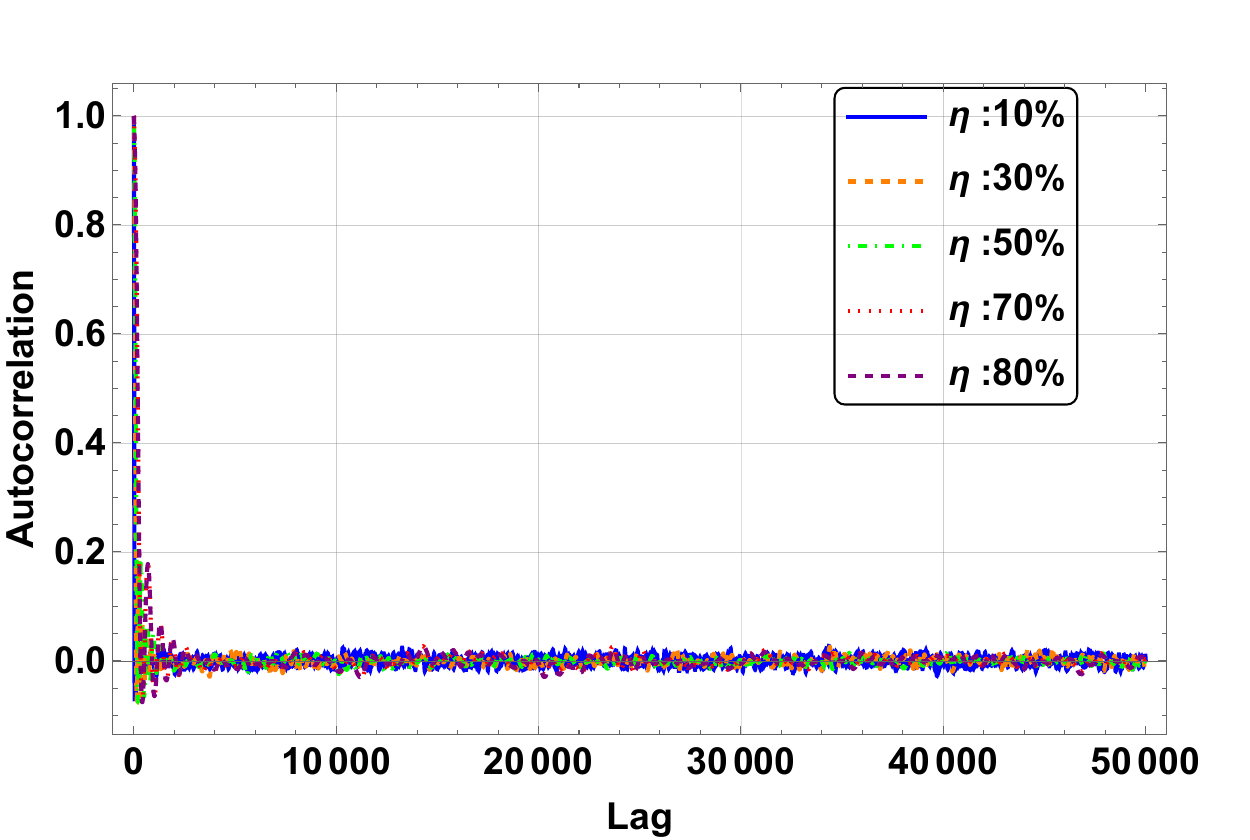}
        \caption{}
    \end{subfigure}
    \caption{(a) Aleatoric Uncertainty $U$ vs Coarse Gaining Percentages (The black dashed line indicates the mean $\overline{U}$). 
             (b) Autocorrelation of coarse-grained data. 
             }
    \label{acf}
\end{figure*}



After training the neural network on the coarse-grained data subsets, we observe that increasing the coarse-graining percentage leads to a corresponding increase in the ensemble-averaged uncertainty $\overline{U}$ and its standard deviation $\sigma_U$ characterizes the variability, as shown in Table \ref{tab:stddev_mse_cg} and Fig. \ref{acf} (a) (the $U$ values). The BNN uncertainty is sensitive to changes introduced by the effective sampling procedure and is in qualitative agreement with previous CV-QRNG analyses showing that source-model entropy estimates $H_{\text{min}}$ are sensitive to finite sampling \cite{PhysRevA.109.052401}. This highlights the importance of accounting for finite-sampling effects in entropy estimation and randomness-extractor design \cite{PhysRevA.109.052401,Zhang2021,PhysRevA.102.012422}.

We also computed the ACF for the coarse-grained datasets. As shown in Fig. \ref{acf} (b), the autocorrelation rapidly decays toward zero across all coarse-graining percentages. The ACF shows only short-lag differences and no clear change in second-order linear correlation over the analyzed lag range; however, this result does not exclude higher-order or longer-range dependencies. In contrast, the BNN metric changes systematically with the effective sampling rate, providing a complementary diagnostic.

\subsection{Studying imperfect precision effects of ADC}
\label{sec:adc2}
Uniform quantization of the observed voltage into an $n$-bit dynamic range is used in CV-QRNG entropy analyses to quantify the effect of finite ADC range and precision $n$ on extractable min-entropy \cite{PhysRevApplied.3.054004,Zhang2021,PhysRevA.102.012422}. The impact of finite ADC precision has also been investigated for pulsed-laser CV-QRNG, where increasing the bit depth $n$ increases $H_{\text{min}}$, with diminishing gains at higher precision \cite{PhysRevA.109.052401}. Here, we investigate analogous effects in a continuous-wave laser setup.

\begin{figure*}[tbp]
    \centering
    \begin{subfigure}[b]{0.59\textwidth}
        \includegraphics[width=\textwidth]{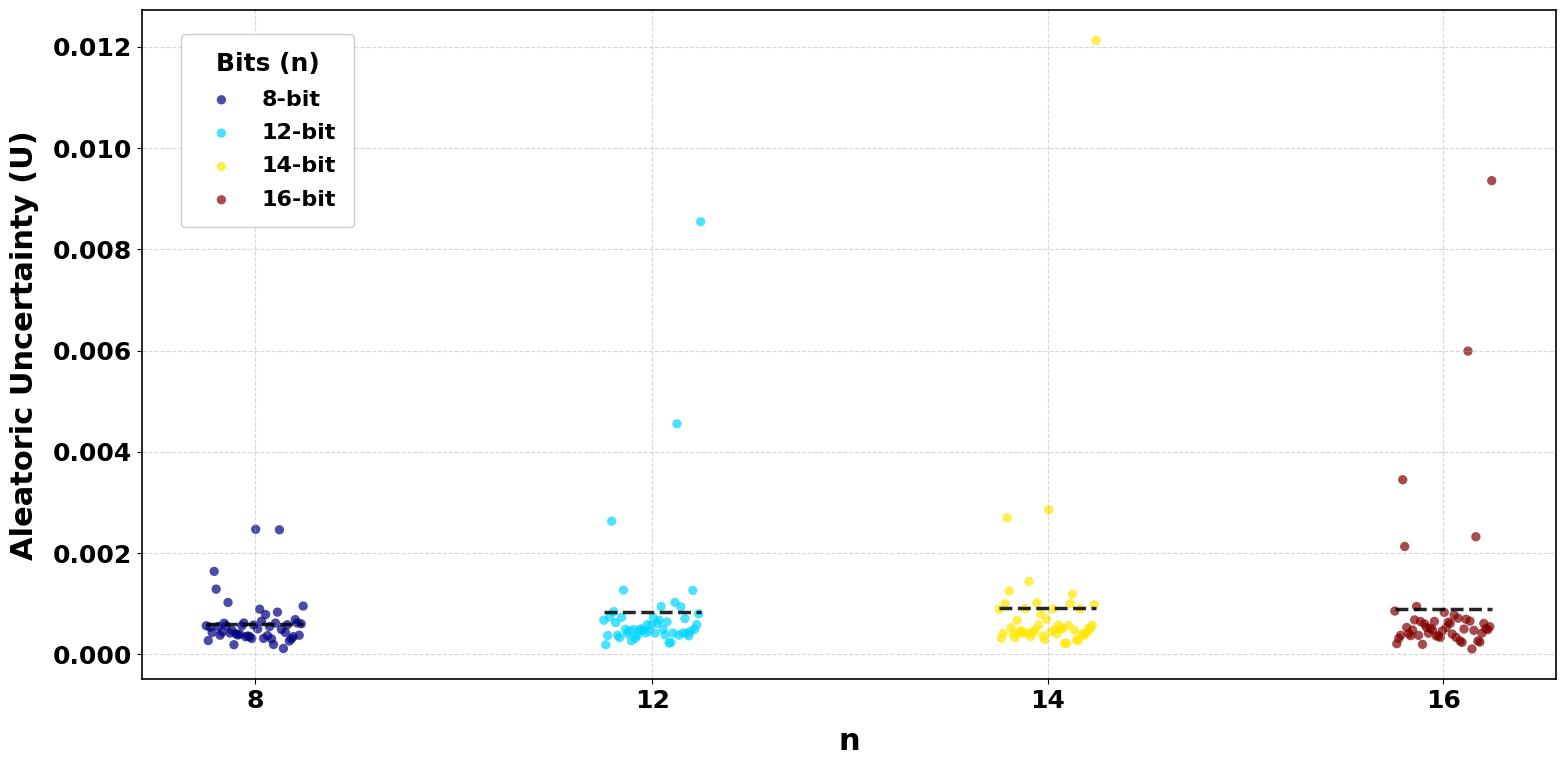}
        \caption{}
    \end{subfigure}
    \hfill
    \begin{subfigure}[b]{0.4\textwidth}
        \includegraphics[width=\textwidth]{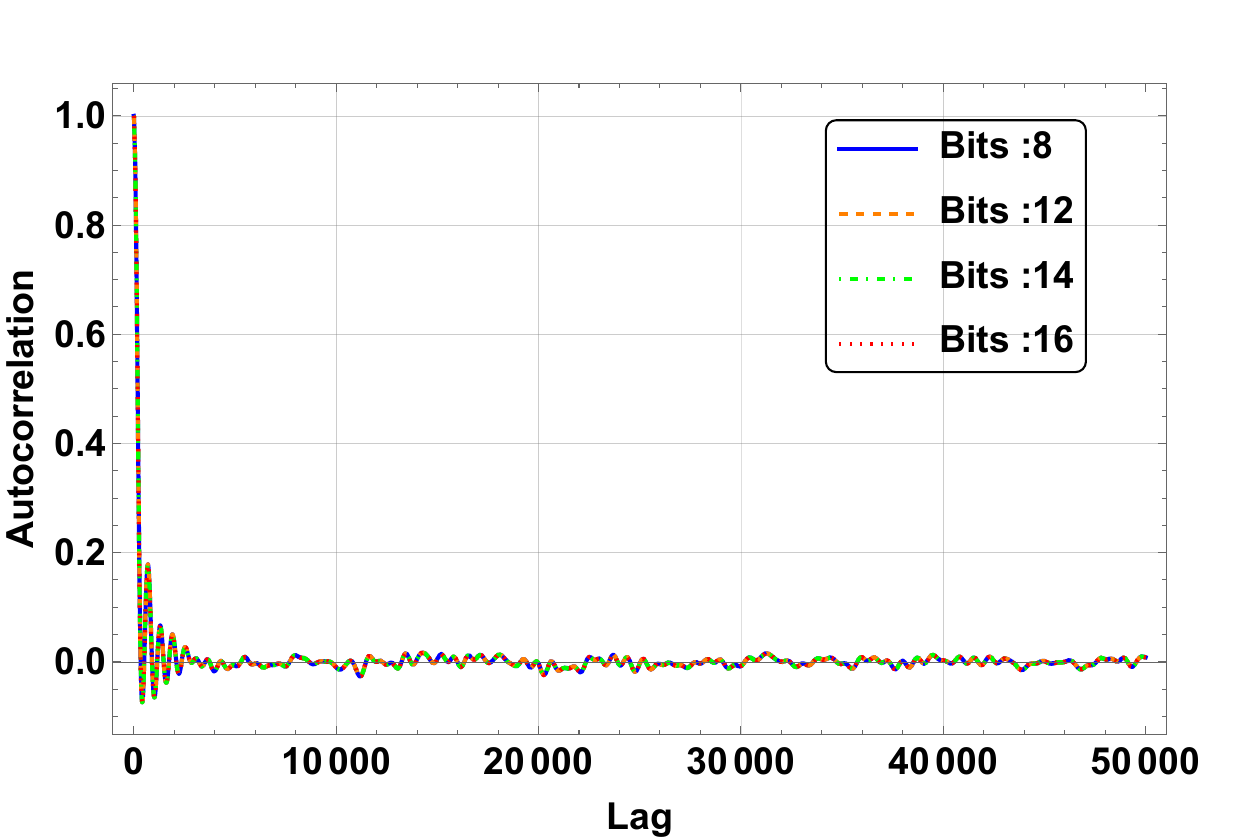}
        \caption{}
    \end{subfigure}
    \caption{(a) Aleatoric Uncertainty $U$ vs ADC sampling precision $n$ (The black dashed line indicates the mean $\overline{U}$). 
             (b) Autocorrelation of voltage data reconstructed at different ADC bit depths. 
             }
    \label{adc}
\end{figure*}

The quantized voltage data (mathematical procedure described in Supplement B) is then processed by the BNN ensemble to obtain the aleatoric uncertainty $U$. As shown in Fig. \ref{adc} (a) and Table \ref{tab:adc_precision}, we observe an increase in the ensemble averaged ($\overline{U}$) and consecutively variability in $\sigma_U$ as the precision increases from 8 to 12 bits and changes only marginally between 12 and 16 bits. This behavior is qualitatively consistent with diminishing influence of ADC precision at $n>10$ \cite{PhysRevA.109.052401}. Previous studies likewise show diminishing gains in $H_{\text{min}}$ at higher ADC precision \cite{PhysRevApplied.3.054004,PhysRevA.102.012422,PhysRevA.109.052401}. Thus, the BNN metric is sensitive to changes in the effective resolution of the measurement apparatus, while this result should not be interpreted as a direct measurement of extractable randomness.


\begin{table}[h!]
\centering
\begin{tabular}{|c|c|c|c|}
\hline
ADC Precision (bits) & $\overline{U}$ & \textbf{$\sigma_{U}$} \\ \hline
8  & $6.086000 \times 10^{-4}$ & $4.697610 \times 10^{-4}$ \\ 
12 & $8.364600 \times 10^{-4}$ & $1.300641 \times 10^{-3}$ \\ 
14 & $9.061600 \times 10^{-4}$ & $1.699422 \times 10^{-3}$ \\ 
16 & $9.039200 \times 10^{-4}$ & $1.542760 \times 10^{-3}$ \\ \hline
\end{tabular}
\caption{Mean Aleatoric uncertainty ($\overline{U}$) and its standard deviation ($\sigma_{U}$) at different ADC sampling precisions.}
\label{tab:adc_precision}
\end{table}


To examine whether ACF could detect these quantization effects, we again computed the ACFs of signals reconstructed at varying bit depths. As illustrated in Fig. \ref{adc} (b), the ACF rapidly decays toward zero for all precision levels, showing no discernible difference between the 8-bit and 16-bit reconstructions. This indicates that the simulated quantization changes are not reflected in the second-order temporal correlation measured by the ACF. In contrast, the change in ($U$) between 8 and 12 bits (Table \ref{tab:adc_precision}) shows that the BNN is sensitive to changes in the the quantized signal that are not apparent in the ACF.

\section{Summary and Discussion}

\label{summary}

We introduced a finite ensemble-BNN framework for monitoring changes in predictive uncertainty from raw analog voltage data for a homodyne-based CV-QRNG. Under the experimental conditions considered, the BNN predicted uncertainty metric changes systematically with LO power, effective sampling, and ADC precision. The LO-power behavior is qualitatively consistent with the expected increase of vacuum-noise contribution relative to classical detector noise and with the corresponding source-model extractable randomness $H_{\mathrm{min}}$ trend. The ADC sampling and precision studies further show that the BNN metric is sensitive to measurement conditions that are not clearly reflected in the autocorrelation function ACF \cite{PhysRevA.87.062327,Kim2023,Zhang2021}.

The BNN uncertainty metric is a BNN model-relative diagnostic and must not be interpreted as an information-theoretic measure of conditional min-entropy $H_{\text{min}}$ or as a proof of quantum unpredictability. In particular, the predictive performance of a finite ensemble BNN does not establish that a more powerful predictor or an adversary with additional classical or quantum side information could not obtain more information about the output. The source-model $H_{\mathrm{min}}$ is conditional on the assumed source, detector-noise, and ADC models \cite{Gabriel2010,PhysRevApplied.3.054004,PhysRevA.102.012422,Zhang2021,PhysRevA.109.052401} and should therefore be distinguished from an adversarial conditional min-entropy bound \cite{WangChau2025}. The BNN metric is consequently intended to complement, rather than replace such physics-based entropy estimation and security-certification methods \cite{Kelsey2015}.

Standard NIST SP 800-22 testing provides an additional statistical assessment of the bit stream generated by the VFQRNG (Supplement C). The 1 GB bit stream passed all 15 statistical tests, with the corresponding $P$-values exceeding the significance threshold $\alpha$. However, passing such tests is not a proof of quantum randomness or information-theoretic security. More exhaustive non-IID, and quantum side-information analyses \cite{NISTSP80090B,Hart2017} would be required for a complete security assessment of higher-order, cross-segment, or longer-range dependencies \cite{Jacak2020,Jóźwiak2024}.

Looking ahead, this neural-network uncertainty analysis could serve as a real-time monitoring layer in practical QRNGs. For instance, the BNN could be implemented on an FPGA for low-latency online monitoring \cite{9586137}. A key advantage of the proposed approach is that it operates directly on the measured analog voltage and can provide an operating-condition-sensitive diagnostic without relying solely on conventional second-order correlation measures.

\section{Code availability}

The code used for training and evaluating the neural networks is publicly available on GitHub \cite{Github2025}.

\bibliography{apssamp}

\pagebreak

\section*{Supplement A: Bayesian Neural network to study randomness in raw voltage data }
\label{sec:Bayesian Neural network to study randomness in raw voltage data}

\subsection{Data Preparation for Timeseries Analysis}
\label{sec:data prep}
The raw voltage time series recorded from the homodyne detector contains approximately $5 \times 10^6$ samples. To ensure computational tractability while maintaining statistical representativeness, the data are analyzed in disjoint subsets. Specifically, each experimental run processes a contiguous segment of $10^4$ voltage samples, and multiple such segments are analyzed independently to assess reproducibility. The measured homodyne-voltage distribution is approximately Gaussian over the operating conditions considered and contains contributions from vacuum fluctuations and classical detector/electronic noise.


Let \textit{N} be the number of data points in the dataset and \textit{L} be the size of the lookback window. Each data segment is standardized prior to training by subtracting its empirical mean and dividing by its empirical standard deviation (zero-mean, unit-variance normalization). We adopt this z-score normalization because it stabilizes training and makes the network's predicted variance independent of the chunk-specific scale during optimization. Normalization is essential for machine learning models because it ensures that all features contribute equally to the training process; without it, features with larger ranges could dominate the model, leading to biased results.

Given the time-series nature of the data, the raw dataset is then segmented into overlapping input-output pairs suitable for neural network training. This is achieved using a sliding window approach: each input consists of a sequence of \textit{L} consecutive, normalized data points $x_i=[v_i,v_{i+1},...,v_{i+L-1}]$, and the corresponding output $y_i=v_{i+L}$ is the next data point that immediately follows this window in the sequence, for $i \in \{1,N-L\}$. This method allows the neural network to learn temporal dependencies by mapping recent history to the subsequent value, which is a standard approach in time series forecasting using neural networks:
\[
\text{Normalized Voltage} = [v_1, v_2, v_3, \dots, v_N].
\]
where $N$ is the total number of normalized data points and for a given look-back window size $L$, the input-output pairs are created as follows:
\[
\begin{aligned}
\text{Input to neural network: } [v_1, v_2, \dots, v_L] \rightarrow \\ \text{Output to neural network: } v_{L+1}
\end{aligned}
\]
\[
\begin{aligned}
\text{Input to neural network: } 
[v_2, v_3, \dots, v_{L+1}] 
\rightarrow \\ \text{Output to neural network: } v_{L+2}
\end{aligned}
\]
This process is repeated for the entire dataset for training and the dataset would be like the following:
\begin{multline}
    \label{eqn2}
\left[
\begin{array}{cccc}
v_1 & v_2 & \cdots & v_L \\
v_2 & v_3 & \cdots & v_{L+1} \\
\vdots & \vdots & \ddots & \vdots \\
v_{N-L} & v_{N-L+1} & \cdots & v_{N-1}
\end{array}
\right]
\quad \longrightarrow \quad
\left[
\begin{array}{c}
v_{L+1} \\
v_{L+2} \\
\vdots \\
v_N
\end{array}
\right]  \\  \quad
\equiv \quad
\{x_i\}\longrightarrow\{y_i\}.
\end{multline}
The resulting dataset is split into training and test sets using an 80:20 ratio. The look-back window \(L\) determines the amount of temporal context provided to the BNN for predicting the next voltage sample. To select an appropriate window size, we considered five values, \(L\in\{10,50,100,200,500\}\), and evaluated the corresponding BNN output aleatoric uncertainty ($U$ defined in next Sec.) for voltage data of LO power 9 mW (based on discussion in Sec. \ref{sec:setup}). The rationale for selecting \(L=200\) is presented at the end of this supplement.

\subsection{ Bayesian Approach to Uncertainty Quantification in CV-QRNGs}

Initially, we employed a deep neural network (DNN) approach to predict $\{y_i\}$ from $\{x_i\}$ (Eq. \ref{eqn2}) and utilized the standard deviation of the Mean Squared Error (MSE) across the ensemble of DNNs as our measure of randomness. Typically, MSE evaluates the neural network's performance for regression tasks, which measures the average of the squared differences between predicted and actual values given by
\begin{equation} \label{eq3}
\text{MSE} = \frac{1}{N_{\text{test}}} \sum_{i=1}^{N_{\text{test}}} ( v_i - \hat{v_i} )^2,
\end{equation}
where $v_i$ is the actual value and $\hat{v_i}$ is the predicted value for $N_{\text{test}}$ predictions. We observed that this ensemble-based metric lacked reproducibility; due to the stochastic nature of weight initialization and optimization in standard deep learning, the MSE variance fluctuated significantly across different training runs on identical datasets. Furthermore, deterministic networks like DNN conflate epistemic uncertainty (uncertainty in model parameters) with aleatoric uncertainty (noise inherent in input dataset) \cite{depeweg2017uncertaintydecompositionbayesianneural}. For a QRNG, only the latter might represent the entropic model-inferred uncertainty of the vacuum fluctuations. Consequently, we transition to a BNN framework here. By placing probability distributions over the network weights, the BNN allows us to mathematically decompose the total predictive variance and isolate the aleatoric uncertainty. This provides a stable, reproducible, and physically interpretable metric for the CV-QRNG, independent of the stochastic variations inherent in the neural network model training process.

\subsubsection{Bayesian Neural Network Architecture}

The BNN employed is a multi-layer perceptron (MLP) designed to operate on the voltage time-series data. Unlike traditional MLPs that learn fixed weight values, this architecture treats weights and biases as random variables and learns their probability distributions. The network receives a univariate input sequence of length $L$ in each instance, a segment of the time-series data (as discussed in Eq. \ref{eqn2}). This input feeds into a series of fully connected (dense) layers that progressively reduce in width. We have used a network with 2 hidden layers, 128 units to 64 units, and finally to an output layer with 2 units, as seen in Fig. \ref{fig:Bayesian_neural_network}. Rectified Linear Unit (ReLU) activation functions are applied after the hidden layers to introduce non-linearity. The output layer is configured to produce two values: the predicted mean and the log-variance of a Gaussian distribution. The log-variance is constrained within a stable range to ensure numerical robustness. The standard deviation is then derived from this log-variance. 

Further, variational inference (VI) is used \cite{Goan_2020,Chai2018UncertaintyEI}, which is an optimization technique that approximates the intractable true posterior distribution of the network's weights by iteratively updating the parameters of a simpler, tractable variational distribution (the "guide"). An automated method is used to construct the variational posterior. This typically involves assuming a simple form for the posterior, such as a diagonal Gaussian distribution for each weight and bias. This means that for each parameter in the network, the guide learns a mean and a standard deviation, effectively approximating each parameter's posterior distribution as a Gaussian. An optimizer, such as Adam, is employed to update the variational parameters of the guide. The training process involves repeatedly sampling weights from the current guide, computing the Evidence Lower Bound (ELBO) loss, and then adjusting the variational parameters to maximize this ELBO. This iterative process refines the guide to better approximate the true posterior.

Once the BNN is trained, the learned guide provides an approximation of the posterior distribution over the weights. To make predictions and quantify uncertainties, it is necessary to sample from this learned posterior. Multiple sets of network weights are sampled from the trained guide. For each sampled set of weights, the network's forward pass is executed using the input data. This yields a set of predictions, including the predicted mean and, crucially, the predicted standard deviation for each input. The predicted standard deviations obtained from these multiple samples are then rescaled back to the original data units. The aleatoric uncertainty for each input is estimated by calculating the mean of the squared predicted standard deviations across all Monte Carlo samples. This average variance captures the inherent noise in the data that the model accounts for. Finally, an overall measure of aleatoric uncertainty is obtained by averaging these individual estimates across all test data points.

This comprehensive approach allows the BNN to generate predictions while providing an estimate of the predictive uncertainty associated with them. The predicted variance representing the BNN model-inferred aleatoric component. Because the homodyne voltage data contain contributions from quantum vacuum fluctuations as well as classical electronic noise, our objective is not to improve point-prediction performance but to characterize the uncertainty associated with predicting the measured signal. Therefore, we focus on the model's predicted variance and use it in this work as the diagnostic metric.

\usetikzlibrary{calc}

\newcommand{\gaussglyph}[1]{%
  \begin{scope}[shift={(#1)}]
    \node[fill=white, draw=red!60!black, rounded corners=1pt,
          minimum width=9mm, minimum height=6mm, inner sep=0pt] {};
    \draw[gray, line width=0.3pt] (-0.38,-0.22) -- (0.38,-0.22);
    \draw[red!70!black, thick, smooth, domain=-1.8:1.8, samples=30, variable=\x]
      plot ({0.2*\x},{-0.22+0.42*exp(-(\x)*(\x))});
  \end{scope}%
}

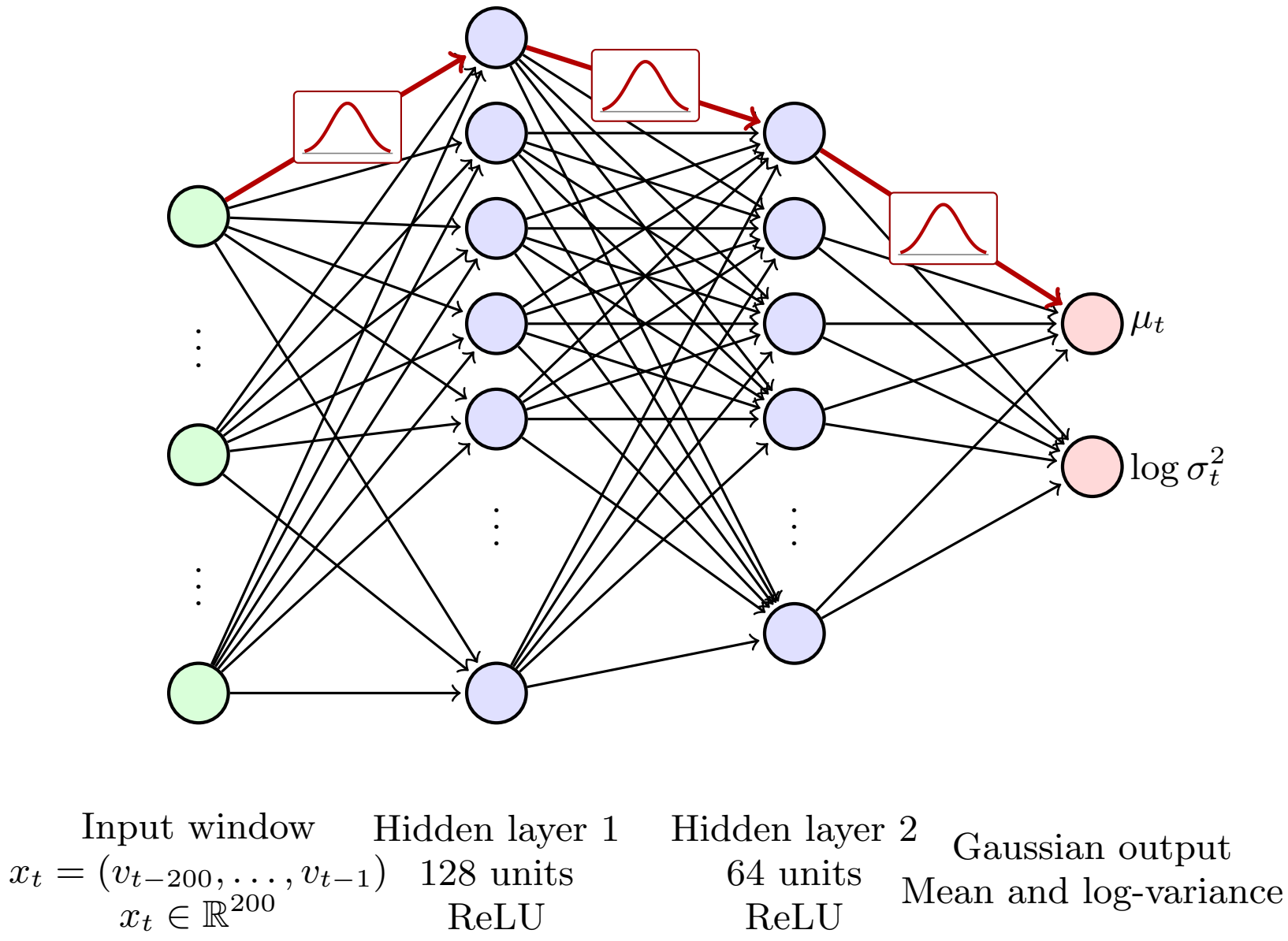
\begin{figure*}[tbp]
\centering
\begin{tikzpicture}[
    neuron/.style={
        circle,
        draw,
        minimum size=5mm,
        inner sep=0pt,
        fill=white,
        thick
    },
    inputneuron/.style={neuron, fill=green!15},
    hiddenneuron/.style={neuron, fill=blue!12},
    outputneuron/.style={neuron, fill=red!15},
    connection/.style={->, shorten >=1pt, semithick},
    layerlabel/.style={align=center, font=\small},
    bayesnote/.style={
        draw,
        rounded corners,
        align=center,
        font=\footnotesize,
        inner sep=5pt
    }
]

\def\layersep{2.5}

\node[inputneuron] (I1) at (0,1.5) {};
\node at (0,0.5) {$\vdots$};
\node[inputneuron] (I2) at (0,-0.5) {};
\node at (0,-1.5) {$\vdots$};
\node[inputneuron] (I3) at (0,-2.5) {};

\node[layerlabel] at (0,-4.0)
{
    Input window\\
    $x_t=(v_{t-200},\ldots,v_{t-1})$\\
    $x_t\in\mathbb{R}^{200}$
};

\foreach \i in {1,...,5} {
    \node[hiddenneuron] (H1\i)
    at (\layersep,{3.0-(\i-1)*0.8}) {};
}
\node at (\layersep,-1.0) {$\vdots$};
\node[hiddenneuron] (H1_final) at (\layersep,-2.5) {};

\node[layerlabel] at (\layersep,-4.0)
{
    Hidden layer 1\\
    128 units\\
    ReLU
};

\foreach \i in {1,...,4} {
    \node[hiddenneuron] (H2\i)
    at ({2*\layersep},{2.2-(\i-1)*0.8}) {};
}
\node at ({2*\layersep},-1.0) {$\vdots$};
\node[hiddenneuron] (H2_final)
at ({2*\layersep},-2.0) {};

\node[layerlabel] at ({2*\layersep},-4.0)
{
    Hidden layer 2\\
    64 units\\
    ReLU
};

\node[outputneuron] (O1) at ({3*\layersep},0.6) {};
\node[outputneuron] (O2) at ({3*\layersep},-0.6) {};

\node[anchor=west,font=\small]
at ({3*\layersep+0.2},0.6) {$\mu_{t}$};

\node[anchor=west,font=\small]
at ({3*\layersep+0.2},-0.6) {$\log\sigma^2_{t}$};

\node[layerlabel] at ({3*\layersep},-4.0)
{
    Gaussian output\\
    Mean and log-variance
};


\foreach \i in {1,2,3} {
    \foreach \j in {1,...,5} {
        \draw[connection] (I\i) -- (H1\j);
    }
    \draw[connection] (I\i) -- (H1_final);
}

\foreach \i in {1,...,5} {
    \foreach \j in {1,...,4} {
        \draw[connection] (H1\i) -- (H2\j);
    }
    \draw[connection] (H1\i) -- (H2_final);
}

\foreach \j in {1,...,4} {
    \draw[connection] (H1_final) -- (H2\j);
}
\draw[connection] (H1_final) -- (H2_final);

\foreach \i in {1,...,4} {
    \draw[connection] (H2\i) -- (O1);
    \draw[connection] (H2\i) -- (O2);
}
\draw[connection] (H2_final) -- (O1);
\draw[connection] (H2_final) -- (O2);

\draw[->, shorten >=1pt, very thick, red!70!black] (I1)  -- (H11);
\draw[->, shorten >=1pt, very thick, red!70!black] (H11) -- (H21);
\draw[->, shorten >=1pt, very thick, red!70!black] (H21) -- (O1);

\gaussglyph{$(I1)!0.5!(H11)$}
\gaussglyph{$(H11)!0.5!(H21)$}
\gaussglyph{$(H21)!0.5!(O1)$}


\end{tikzpicture}
\caption{Architecture of one Bayesian neural network (BNN) member used
for next-sample prediction of the homodyne voltage. The input
consists of $L=200$ consecutive voltage samples. Two fully connected
hidden layers contain 128 and 64 ReLU units, respectively. The
output layer predicts the mean $\mu_{t}$ and log-variance
$\log\sigma^2_{t}$ of a Gaussian likelihood. Network weights and
biases are treated probabilistically using variational inference,
which approximates their posterior distribution $q_{\phi}(w)$. The drawing shows representative neurons; the red
edges and Gaussian glyphs mark example weights whose values are drawn from learned
distributions, and the labels give the full layer dimensions. Every weight and bias is a random variable.}
\end{figure*}

\subsubsection{Bayesian Neural Networks and Aleatoric Uncertainty Estimation}

Given a dataset $D$ such as in Eq. \ref{eqn2}, the BNN aims to learn the posterior distribution $p(w|D)$ over its network weights $w$. According to Bayes' rule, this posterior is given by \cite{Goan_2020,Chai2018UncertaintyEI}:

$$p(w|D) = \frac{p(D|w)p(w)}{p(D)}$$

where
\begin{itemize}
    \item $p(D|w)$ is the likelihood of the data given the weights. For regression task like ours, this is often modeled as a Gaussian likelihood, $p(y|w, x) = \mathcal{N}(y; \mu', \sigma')$ (normal distribution of $y$), where $\mu'$ and $\sigma'$ (their parameters) are predicted by the neural network from the input $x$.
    \item $p(w)$ is the prior distribution over the weights $w$, reflecting our initial beliefs about the weights before observing any data.
    \item $p(D)$ is the evidence, a normalizing constant defined as $p(D)=\int p(D|w) p(w) dw$.
\end{itemize}

  In BNNs, the true posterior distribution $p(w|D)$ over the network's weights is often analytically intractable to compute. As mentioned, VI is employed to overcome this. VI seeks to approximate the true posterior $p(w|D)$ with a simpler, more tractable distribution $q(w|\phi)$, known as the variational posterior or guide. Here, $\phi$ represents the variational parameters that define this approximating distribution \cite{NIPS2011_7eb3c8be}. The objective of VI is to minimize the Kullback-Leibler (KL) divergence between the variational posterior $q(w|\phi)$ and the true posterior $p(w|D)$. However, directly minimizing this KL divergence is also intractable because it requires computing the evidence $p(D)$. Instead, VI focuses on maximizing a lower bound on the marginal likelihood $p(D)$, which is known as the ELBO.

Further, in our setting, the BNN is designed to output both the mean ($\mu$) and the variance ($\sigma^2$) of a Gaussian distribution for each prediction. The aleatoric uncertainty is then directly captured by this predicted variance. To estimate the total aleatoric uncertainty $U$, we find the expectation of the predicted variance over the posterior distribution of the weights \cite{Chai2018UncertaintyEI} :
\begin{equation} \label{eqn4}
    U \approx \frac{1}{N_{\text{test}}} \sum_{i=1}^{N_{\text{test}}} \frac{1}{T} \sum_{t=1}^{T} \hat{\sigma}_{i,t}^2,
\end{equation}
where
\begin{itemize}
    \item $T=25$ is the number of Monte Carlo samples, which was found to be sufficient to stabilise the estimate of $U$, while maintaining computational efficiency.
    \item $\hat{\sigma}_{i,t}^2$ is the predicted variance for the $t^{th}$ sampled set of weights of $i^{th}$ test sample.
\end{itemize}

Post-training, the variational parameters defining the approximate posterior $q(w|\phi)$ are fixed. To compute $U$ for a specific input sequence \{$x_i$\}, we perform $T$ stochastic forward passes (Monte Carlo sampling). In each pass $t$, a distinct set of weights $w_t$ is sampled from the learned distribution $q(w|\phi)$. The network processes the same input \{$x_i$\} using these sampled weights to output a predicted variance $\hat{\sigma}^2_{i,t}(x_i, w_t)$. This procedure effectively integrates over the uncertainty in the model weights. 

\subsection{Model Evaluation process}

As described in the previous section, aleatoric uncertainty ($U$) from Eq. \ref{eqn4} evaluates the neural network's performance. The voltage stream acquired from the balanced homodyne detector is partitioned into 50 mutually exclusive subsets. Each subset is delivered to a dedicated BNN member of an ensemble $E = \{m_{1},m_{2},\cdots,m_{j},\cdots,m_{50}\}$, and every network is trained independently. After convergence, the aleatoric uncertainty recorded for the $j^{th}$ model is collected as $U = \{U_{1},U_{2},\cdots,U_{j},\cdots,U_{50}\}$. We then compute the mean as

\begin{equation}
     \overline{U} =  \frac{1}{50} \sum_{j=1}^{50} \left( {U}_j \right), 
\end{equation}
standard deviation as
\begin{equation}
     \sigma_U = \sqrt{ \frac{1}{49} \sum_{j=1}^{50} \left( {U}_j - \overline{U} \right)^2 },
\end{equation}
and treat these as an empirical indicator and diagnostic metric.  Since the aleatoric uncertainty ($U_{j}$) serves as a quantitative measure of how well each model within the ensemble learns the data, examining the standard deviation of $U$ values across all models provides valuable insight into the consistency of the ensemble’s predictions, similar to MSE from Eq. \ref{eq3} (For DNN). A high mean $\overline{U}$ and variability of standard deviation $(\sigma_{U})$ of $U$ indicates that there is substantial variability in the $U$ from one model to another. This variability suggests that the models are not consistently able to learn or generalize patterns from the data, and may, in effect, be resorting to random guessing in their predictions. If the standard deviation is low, it implies that the models are performing similarly, potentially indicating that the data contains learnable, deterministic structure. Conversely, a high standard deviation points to the presence of randomness that the models are unable to capture, reflecting the stochastic nature of the dataset. 

\subsubsection{Sensitivity to look-back window size $L$}
\label{subsec:lookback_sensitivity}

\begin{figure*}[tbp]
    \centering
\includegraphics[width=0.9\textwidth]{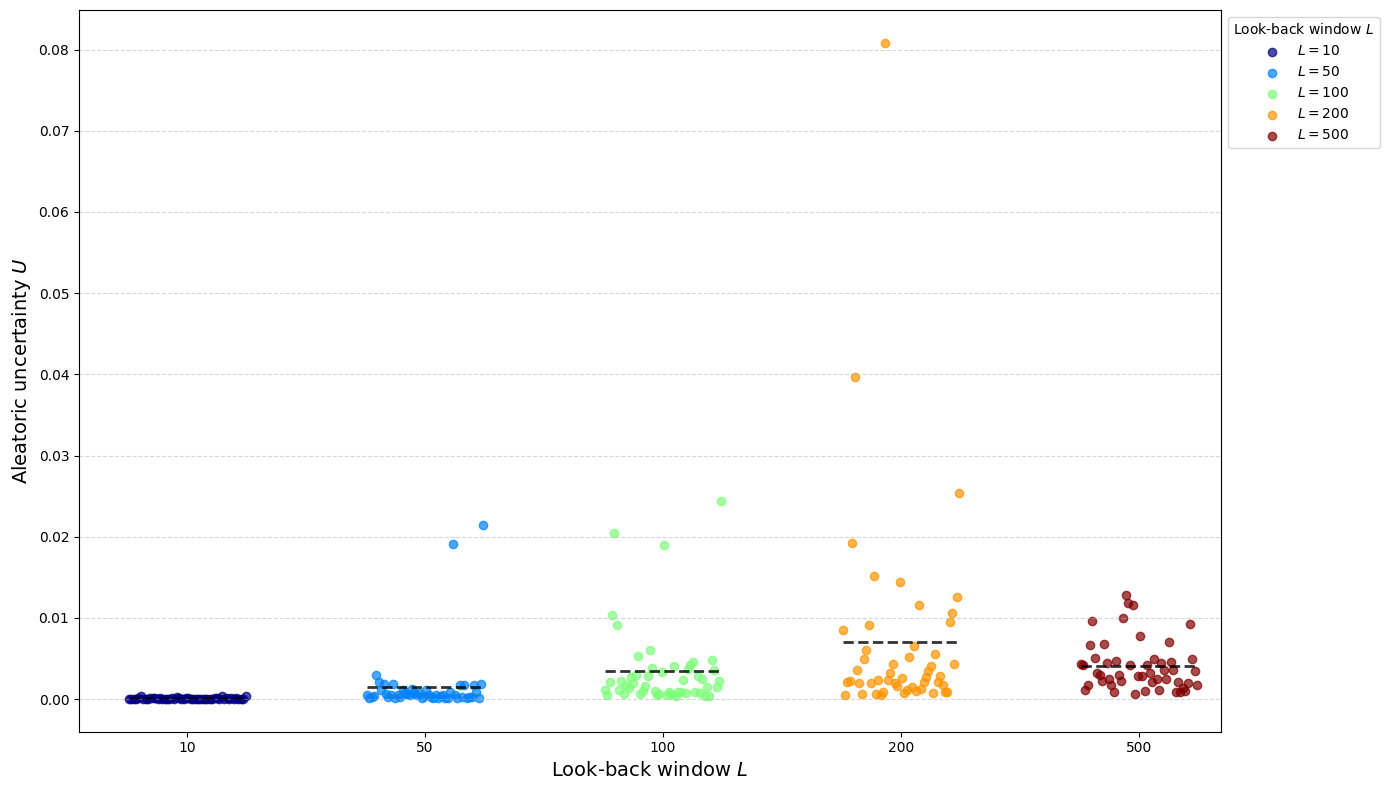}
    \caption{Aleatoric Uncertainty $U$ vs Look-back window $L$ (The black dashed line indicates the mean $\overline{U}$).}
    \label{fig:lookback}
\end{figure*}

To investigate the influence of temporal context on the proposed aleatoric uncertainty metric $U$, we evaluated five look-back window sizes, \(L\in\{10,50,100,200,500\}\). The resulting ensemble mean $\overline{U}$ and standard deviation $(\sigma_{U})$ of the predicted aleatoric uncertainty $U$ are summarized in Table~\ref{tab:lookback_sensitivity} and their values are shown in Fig. \ref{fig:lookback}

\begin{table}[h!]
\centering
\begin{tabular}{|c|c|c|}
\hline
Look-back ($L$) & $\overline{U}$ & \textbf{$\sigma_{U}$} \\ \hline
10  & $8.600 \times 10^{-5}$ & $8.600 \times 10^{-5}$ \\
50  & $1.499 \times 10^{-3}$ & $3.895 \times 10^{-3}$ \\
100 & $3.480 \times 10^{-3}$ & $4.988 \times 10^{-3}$ \\
200 & $6.987 \times 10^{-3}$ & $12.727 \times 10^{-3}$ \\
500 & $4.080 \times 10^{-3}$ & $3.040 \times 10^{-3}$ \\ \hline
\end{tabular}
\caption{Mean aleatoric uncertainty ($\overline{U}$) and its standard deviation ($\sigma_{U}$) at different Look-back values ($L$).}
\label{tab:lookback_sensitivity}
\end{table}

The mean aleatoric uncertainty $\overline{U}$ and its standard deviation $\sigma_U$ both increase as the look-back window increases from $L=10$ to $L=200$, reaching their highest observed values at $L=200$. For $L=500$, both quantities decrease. Thus, $L=200$ yields the highest observed mean aleatoric uncertainty $\overline{U}$ and the largest standard deviation $\sigma_U$. These results indicate that increasing the look-back window does not necessarily lead to a higher value of the proposed uncertainty metric $U$. We therefore selected $L=200$ for the subsequent analyses because it yields the highest observed mean aleatoric uncertainty among the tested, while avoiding the additional computational burden associated with the larger windows. The comparatively large standard deviation at $L=200$ also indicates substantial variability.

    \section*{Supplement B: Modeling of autocorrelation function (ACF), imperfect sampling effects of ADC and imperfect precision effects of ADC}

\label{sec:math}

\subsection{ACF Modeling}

The ACF quantifies the linear dependence between the signal and a time-shifted version of itself; we compute it as following:

\begin{itemize}
    \item The global arithmetic mean of the original voltage sequence $\{v_j\}$ is computed to center the data:
    \begin{equation}
        \bar{v} = \frac{1}{M} \sum_{j=0}^{M-1} v_j.
    \end{equation}
    
    \item For a specific time lag $k$, the normalized ACF $\rho_k$ is obtained by dividing the autocovariance at lag $k$ by the variance (autocovariance at lag 0). This normalizes the result to the range $[-1, 1]$:
    \begin{equation}
        \rho_k = \frac{\sum_{j=0}^{M-k-1} (v_j - \bar{v})(v_{j+k} - \bar{v})}{\sum_{j=0}^{M-1} (v_j - \bar{v})^2}.
    \end{equation}
\end{itemize}

Given that the voltage data was recorded using an oscilloscope with a sampling rate of 40 GHz, the temporal resolution (time step) between data points is $\Delta t = 1 / (40 \times 10^9) \text{ s} = 25 \text{ ps}$. Consequently, in our analysis we calculate the autocorrelation up to a lag of $k=50,000$, this corresponds to a total physical time delay of $\tau = 50,000 \times 25 \text{ ps} = 1.25 \mu\text{s}$.

\subsection{Mathematical modeling of imperfect sampling effects of ADC}

Mathematically, we model imperfect sampling effects of ADC as a discrete block-averaging operation defined by the following steps:

\begin{itemize}
    \item Let the original digitized voltage time-series be denoted by 
    \begin{equation}
       \mathcal{V} = \{v_i\}_{i=0}^{N-1}, 
    \end{equation}
     where $N$ is the total number of raw data points.
    
    \item For a defined retention percentage $\eta \in (0, 1]$, the target number of samples ($M$) is calculated as the nearest integer:
    \begin{equation}
        M = \text{round}(\eta N).
    \end{equation}
    
    \item This imposes an effective sampling window ($S$) determined by integer division:
    \begin{equation}
        S = \lfloor N/M \rfloor.
    \end{equation}
    
    \item The coarse-grained voltage sequence $\mathcal{V}' = \{v'_j\}_{j=0}^{M-1}$ is constructed by averaging the original signal over non-overlapping, contiguous windows of size $S$. The $j$-th downsampled value is given by:
    \begin{equation}
        v'_j = \frac{1}{S} \sum_{k=0}^{S-1} v_{(j \cdot S + k)}.
    \end{equation}
\end{itemize}

\subsection{Mathematical modeling of imperfect precision effects of ADC}

To simulate imperfect precision effects of ADC, we apply uniform quantization to the raw voltage time series. This process maps a continuous-valued voltage signal to a discrete set of approximation levels $\mathcal{V}\rightarrow\mathcal{V}'$ defined by the simulated ADC's precision, similar to the previous section. 

\begin{itemize}
    \item Let $v_{\min}$ and $v_{\max}$ denote the global minimum and maximum values of the acquired voltage signal $\mathcal{V}$. For a simulated $n$-bit ADC, the dynamic range $v_{\max} - v_{\min}$ is partitioned into $N = 2^n$ discrete quantization intervals.
    \item The uniform quantization step size, or bin width $\Delta$, is defined as:
\begin{equation}
    \Delta = \frac{v_{\max} - v_{\min}}{N}
\end{equation}
\item Each instantaneous voltage sample $v_i \in \mathcal{V}$ is mapped to a discrete bin index $k(v_i) \in \{0, 1, \dots, N-1\}$. Since the signal is bounded by $[v_{\min}, v_{\max}]$, the index is determined by the floor function, constrained by the upper limit $N-1$:

\begin{equation}
    k(v_i) = \min \left( \left\lfloor \frac{v_i - v_{\min}}{\Delta} \right\rfloor, N - 1 \right)
\end{equation}
\item To maintain a continuous domain representation suitable for the BNN, the signal is reconstructed into $\mathcal{V}'$ by mapping each index $k$ to the geometric midpoint of its corresponding interval:

\begin{equation}
    v'_i = v_{\min} + \left(k(v_i) + \frac{1}{2}\right) \Delta
\end{equation}
\end{itemize} 

\section*{SUPPLEMENT C: National Institute of Standards and Technology statistical tests}
\label{sec:NIST}

\begin{figure}[H]
    \centering
    \includegraphics[width=8cm]{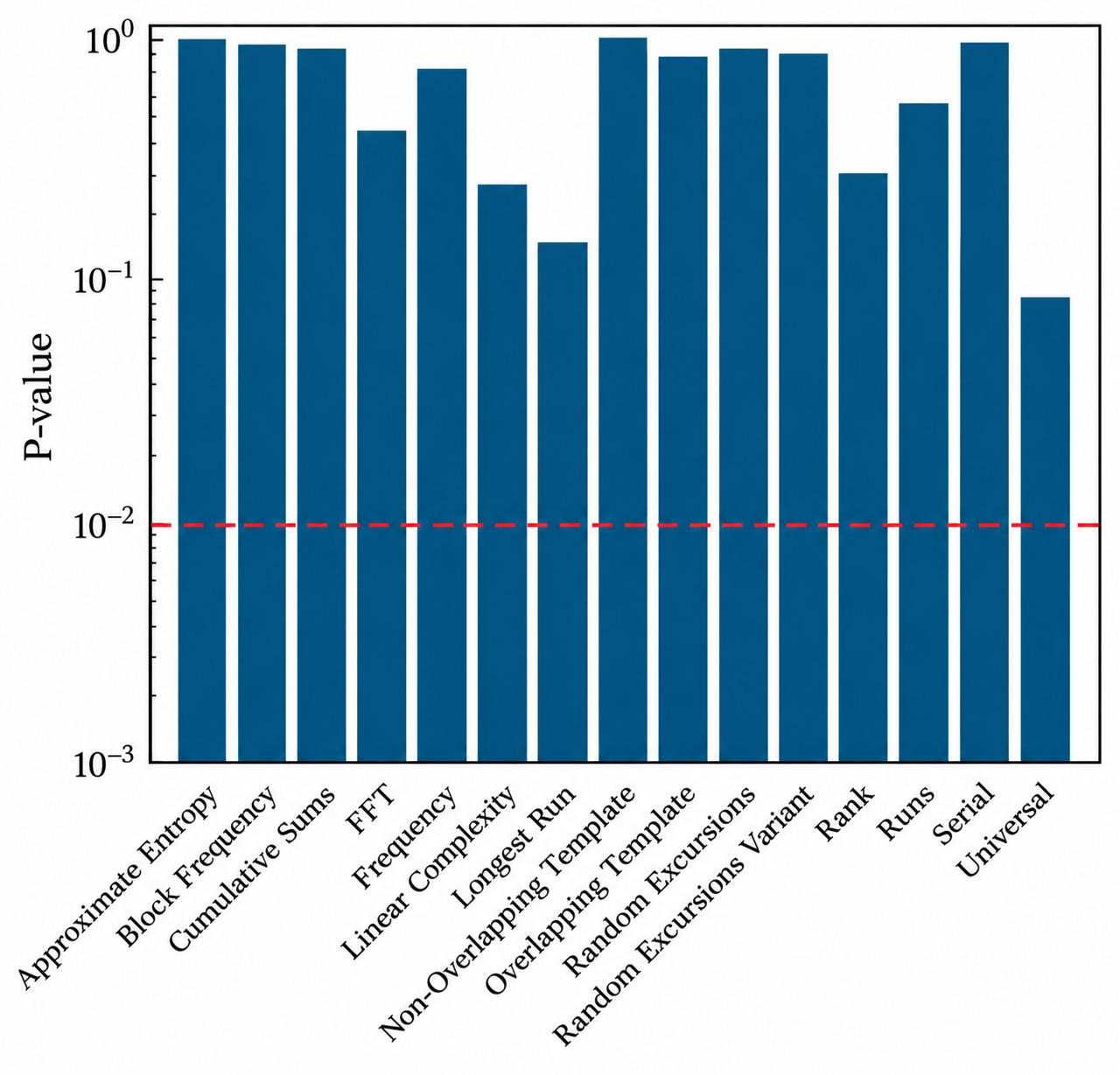}
    \caption{P-values obtained from the statistical tests of the NIST (National Institute of Standards and Technology) test suite are presented. According to the NIST SP 800-22 criteria, a test is considered to pass when the corresponding P-value exceeds the significance threshold of 0.01.}
    \label{fig:enter-label}
\end{figure}

The quality of the generated randomness was evaluated using 1 GB of recorded bit data using 14-bit ADC \cite{Shrivastava2025} and the 15 statistical tests specified in the NIST SP 800-22 test suite \cite{lawrence}. Prior to performing the tests, a significance level $\alpha$ is selected, which denotes the probability of rejecting a genuinely random sequence. A commonly adopted value in cryptographic applications is $\alpha=0.01$. For each statistical test, the resulting P-value quantifies the consistency of the observed sequence with the null hypothesis of randomness. P-values close to unity indicate strong consistency with random behavior, whereas values approaching zero provide evidence against the null hypothesis. Accordingly, the null hypothesis is accepted when P-value $\geq \alpha$, while a P-value below $\alpha$ leads to its rejection. As illustrated in Fig. 6, the P-values obtained from all 15 NIST statistical tests remain above the chosen threshold of $\alpha=0.01$, demonstrating that the recorded sequence passes the complete set of randomness tests.

\end{document}